# Direct observation of nodeless superconductivity and phonon modes in electron-doped copper oxide $Sr_{1-x}Nd_xCuO_2$

Jia-Qi Fan[1], Xue-Qing Yu[1], Fang-Jun Cheng[1], Heng Wang[1], Ruifeng Wang[1], Xiaobing Ma[1], Xiao-Peng Hu[1], Ding Zhang[1,2,3,4], Xu-Cun Ma[1,2 †], Qi-Kun Xue[1,2,3,5 †], Can-Li Song[1,2 †]

[1]State Key Laboratory of Low-Dimensional Quantum Physics, Department of Physics, Tsinghua University, Beijing 100084, China

[2]Frontier Science Center for Quantum Information, Beijing 100084, China

[3]Beijing Academy of Quantum Information Sciences, Beijing 100193, China

[4]RIKEN Center for Emergent Matter Science (CEMS), Wako, Saitama 351-0198, Japan

[5]Southern University of Science and Technology, Shenzhen 518055, China

**Abstract**

The microscopic understanding of high-temperature superconductivity in cuprates has been hindered by the apparent complexity of crystal structures in these materials. We used scanning tunneling microscopy and spectroscopy to study an electron-doped copper oxide compound $Sr_{1-x}Nd_xCuO_2$ that has only bare cations separating the $CuO_2$ planes and thus the simplest infinite-layer structure among all cuprate superconductors. Tunneling conductance spectra of the major $CuO_2$ planes in the superconducting state revealed direct evidence for a nodeless pairing gap, regardless of variation of its magnitude with the local doping of trivalent neodymium. Furthermore, three distinct bosonic modes are observed as multiple peak-dip-hump features outside the superconducting gaps and their respective energies depend little on the spatially varying gaps. Along with the bosonic modes with energies identical to those of the external, bending and stretching phonons of copper oxides, our findings indicate their origin from lattice vibrations rather than spin excitations.

†*To whom correspondence should be addressed. Email: clsong07@mail.tsinghua.edu.cn, xucunma@mail.tsinghua.edu.cn, qkxue@mail.tsinghua.edu.cn*

**Introduction**

Despite more than three decades of intensive research, it remains a mystery how high-temperature ($T_c$) superconductivity works in a family of ceramic materials known as cuprates [1,2]. In pursuit of its microscopic mechanism, two fundamental prerequisites are needed to identify the superconducting energy gap ($\Delta$) function and bosonic glue (e.g. lattice vibrations and spin excitations) to pair electrons of the copper oxide ($CuO_2$) planes. In theory [3], the bosonic excitation (mode) often exhibits itself, via a strong coupling to the paired electrons, in the low-lying quasiparticle states at energy $E = \Delta + \Omega$ ($\Omega$ is the boson energy). This shows great promise for simultaneously measuring the $\Delta$ and $\Omega$ by tunneling spectroscopy, which has unequivocally established the phonon-mediated *s*-wave pairing state in conventional superconductors [4]. For cuprate superconductors, however, no consensus exists on both the superconducting gap symmetry and the pairing glue [2,5-12]. One explanation could be that previous tunneling data from surface-sensitive scanning tunneling microscopy (STM) have been mostly measured on various charge reservoir layers [8,10-12], where a nodal gap behavior probably associated with charge density wave often occurs [2,13]. Here we report high-resolution STM study of an electron-doped cuprate compound $Sr_{1-x}Nd_xCuO_2$ (SNCO, $x \sim 0.100$), directly reveal nodeless superconductivity and three distinct bosonic modes on the $CuO_2$ planes. Our analysis of the bosonic mode energies, which depend little on the spatially varying $\Delta$, supports a lattice vibrational origin of the modes consistent with external, bending and stretching phonons of the copper oxides.

Infinite-layer $SrCuO_2$ is one of the structurally simplest cuprate parent compounds that comprises the essential $CuO_2$ planes separated only by strontium atoms. Partial substitution of divalent strontium ($Sr^{2+}$) by trivalent neodymium ($Nd^{3+}$) ions leads to electron doping and superconductivity with a record electron-doped cuprate transition temperature $T_c$ of 40 K [14]. Most importantly, the single-crystalline SNCO epitaxial films with well-controlled doping level $x$, grown on $SrTiO_3(001)$ substrates with an oxide molecular beam epitaxy (MBE), exhibit a rare surface termination of the essential $CuO_2$ planes [15-17]. Tunneling spectra of the electron-doped infinite-layer cuprates, which have so far been largely unexplored as compared to their hole-doped counterparts [8], pose considerable challenges and opportunities. The challenges are to clarify whether the electron-doped cuprates and direct measurements of the major

CuO$_2$ planes are fundamentally different from or analogous to their hole-doped counterparts and those of the charge reservoir planes, respectively, whereas the critical opportunity is that addressing these issues might help greatly in finding the culprit of high-$T_c$ superconductivity in the copper oxide superconductors.

## Results

### Temperature-dependent resistivity in Sr$_{1-x}$Nd$_x$CuO$_2$

Figure 1a plots the temperature dependence of electrical resistivity of SNCO with varying Nd doping concentration $x$. An insulator-superconductor transition triggered by Nd$^{3+}$ dopants becomes evident at $x > 0.080$. The superconducting phase is further unambiguously confirmed by applying an external magnetic field to the $x = 0.107$ SNCO sample in Fig. 1b. As anticipated, the electrical resistivity at low temperatures is elevated with increasing field until a complete suppression of superconductivity at 8 T. It is worth noting that the resistivity does not drop down to zero below $T_c$. Instead, it exhibits an upturn behavior, which is later revealed by site-resolved tunneling spectroscopy to arise from nanoscale electronic phase separation between the superconductivity and underdoped Mott insulating state. Anyhow, the observed $T_c$ onset (marked by the black arrows in Fig. 1a) up to 30 K turns out to be higher than those previously reported in the epitaxial SNCO films on the SrTiO$_3$ substrates [18].

### Spectroscopic evidence of nodeless superconductivity and phonon modes

As demonstrated before [15-17], the heteroepitaxy of SNCO films on SrTiO$_3$ proceeds in a typical layer-by-layer mode. Figure 1c shows a constant-current STM topographic image that displays atomically flat and defect-free copper oxide surface in one SNCO sample of $x \sim 0.100$. The adjacent Cu atoms are spaced $\sim 0.39$ nm apart, which agrees with the previous reports [14-17]. In Fig. 1d, we show the energy-resolved tunneling conductance (dI/dV) spectra, being proportional to the quasiparticle density of states (DOS), directly on the CuO$_2$ plane. At 4.8 K, the spectral weight is completely removed over a finite energy range around the Fermi level ($E_F$), and instead considerable DOS piles up at two $E_F$-symmetric gap edges of about $\pm 19$ meV. These characteristics, hallmarks of fully-gapped superconductivity, suggest no gap node in the superconducting gap function of SNCO on the Fermi surface. At elevated temperatures, the superconducting gap is progressively smeared out and vanishes at 78 K (see the red curve). Note that, albeit weak (green curve), the gap survives above the observed $T_c$ maximum of $\sim$

30 K as shown in the electrical transport measurements in Fig. 1a, which we here ascribe to a spatial inhomogeneity of $T_c$ inside the SNCO films.

Careful measurements on various samples and superconducting regions indicate that the nodeless electron pairing occurs universally on the $CuO_2$ planes of SNCO, irrespective of the spatial inhomogeneity in $\Delta$ (Figs. S1 and S2 in the online supplementary material). Such finding turns out to be consistent with previous tunneling and angle-resolved photoemission studies of a sister compound $Sr_{0.9}La_{0.1}CuO_2$ [19,20]. Furthermore, multiple peak-dip-hump fine structures develop frequently (> 75%) outside the superconducting gaps, which are pairwise centered at $E_F$ and smear out at elevated temperatures (Fig. 1d). These traits were commonly interpreted as the signatures of bosonic excitations in superconductors [4,10-12]. Tunneling spectra on the bosonic mode energy $\Omega$ and its correlation with $\Delta$ can potentially distinguish between candidates for the pairing glue [10-12,21]. Inserted in Fig. 2a are one representative superconducting spectrum (black curve) and its derivative ($d^2I/dV^2$, red curve), only showing the empty states. By taking the maxima in the second derivative of conductance $d^2I/dV^2$ as estimates of the energies $E = \Delta + \Omega$, three bosonic modes with energies at $\Omega_{1,2,3}$ are extracted. A statistical estimate of $\Omega$ from all measured superconducting dI/dV spectra in both empty and occupied states yields thousands of independent observables, whose histograms are plotted in Fig. 2a. The average mode energies are measured to be of $\Omega_1$ = 20 ± 3 meV, $\Omega_2$ = 45 ± 4 meV, and $\Omega_3$ = 72 ± 3 meV, respectively. Evidently, $\Omega_2$ and $\Omega_3$ are not multiples of $\Omega_1$ (Fig. S3 in the online supplementary material), which excludes the possibility that they are caused by a harmonic multi-boson excitation of the same mode $\Omega_1$. This appears to be in good agreement with the intensity differences of $\Omega_{1,2,3}$ (Figs. 1d and S1 in the online supplementary material). For some spectra, the bosonic modes $\Omega_2$ and $\Omega_3$ are too faint to be easily read out, leading to their relatively lower probabilities than $\Omega_1$ in the histograms of $\Omega_{1,2,3}$ (Fig. 2a).

The correlations between the spatially-resolved $\Omega_{1,2,3}$ and $\Delta$ are plotted in Figs. 2b and 2c. Despite a substantial spatial variation in $\Delta$ (Fig. S2 in the online supplementary material), the bosonic mode energies $\Omega_{1,2,3}$ alter little and the local ratio of $\Omega_{1,2,3}$ to $2\Delta$ ($\Omega_{1,2,3}/2\Delta$) exceeds unity for small $\Delta$. Both findings run counter to the scenario of spin excitations, whose energies are generally dependent on $\Delta$ [11] and remain below the pair-breaking energy, to wit, $\Omega/2\Delta$ < 1 [21]. By contrast, since energies of lattice vibrations change little with the doping level (i.e.

$\Delta$), they are natural candidates for the three bosonic excitations observed. Actually, the $\Omega_{1,2,3}$ show incredible coincidences with the external (~ 20 meV), bending (~ 45 meV) and stretching (~ 72 meV) phonon mode energies, which were measured independently by optics [22] and Raman [23] in bulk $SrCuO_2$. It should be stressed that this concurrent observation of the three key phonon modes from one spectrum is rather challenging in cuprate superconductors. We attribute this unprecedented success as a new rare measure of the major $CuO_2$ planes.

**Nanoscale electronic phase separation**

To eliminate any possible artifacts in our measurements, we collected spatially resolved tunneling dI/dV spectra over many regions of the samples and checked the tunneling junction quality. A representative set of dI/dV spectra in the superconducting region exhibit superior robustness of the nodeless pairing, coherence peaks as well as peak-dip-hump line shapes (Fig. 3a). Figure 3b compares a series of site-specific dI/dV spectra taken as a function of increasing tip-to-sample distance (from top to bottom). The full superconducting gap remains essentially unchanged at any tunneling current as anticipated for an ideal vacuum tunneling. It was also found that a fraction of superconducting gaps display pronounced coherence peaks and could be fitted by the Dynes model using a single *s*-wave gap function [24], as exemplified in Fig. 3c. A minor discrepancy occurs between measured (black circles) and fitted (red curve) curves in the superconducting gap and suggests excess subgap DOS, whose origin merits further study. Figure 3d exhibits another region in which a set of spatially dependent dI/dV spectra (Fig. 3e) were taken along the red arrow. Note that the atomically-resolved STM topography of Fig. 3d was acquired at -1.0 V, just around the charge-transfer band (CTB) onset of $CuO_2$ in the electron-doped SNCO films [16,17]. The STM contrast mainly has an electronic origin (Fig. S4 in the online supplementary material). Domains mapped as bright correspond to the regions with relatively heavier neodymium dopants and thereby more emergent in-gap states (IGS), and vice versa [16,25]. A local measurement of $E_F$ relative to the midgap energy $E_i$ (i.e. the center of charge transfer gap) of $CuO_2$ [16], by exploring the spatial dependence of wider-energy-ranged dI/dV spectra in Supplementary Fig. S5, convincingly supports this claim. As verified in the bottom panel of Fig. 3d, the $E_F - E_i$ value, a good indicator of electron doping level of neodymium, appears to be larger in the bright regions than that in the dark ones. Meanwhile, a crossover from the full superconducting gaps to gapless or somewhat insulating tunneling spectra is apparent as the STM tip moves from the bright regions to dark ones (Fig. 3e). Such direct imaging of the nanoscale electronic phase separation, which proves a generic characteristic of the SNCO epitaxial films (Fig. S2 in the online supplementary material) and has been extensively documented in other copper oxide superconductors [26,27], offers a straightforward account for the unusual electrical resistivity behavior in Figs. 1a and 1b.

In order to cast more light on the superconductivity of $CuO_2$ at the nanoscale, we further mapped dI/dV spectra in both wide (from -1.5 to 1.5 V) and narrow (from -200 mV to 200 mV) voltage ranges, from which the local doping level and Δ can be readily extracted and compared in the same field of view. Data from such maps are shown in Fig. 4a. Here the spatial IGS are estimated by integrating the spectral weights within the charge-transfer gap of the $CuO_2$ plane, while the Δ is defined as half the separation between the two $E_F$-symmetric superconducting coherence peaks. A direct visual comparison between the corresponding maps in Fig. 4a, and their cross-correlation in Fig. 4b, reveal that the larger $E_F$ - $E_i$ is correlated with topographically bright regions of populated IGS on a short length scale of ∼ 3 nm. By contrast, the correlation between the $E_F - E_i$ value and Δ is too small to draw any unbiased opinion. A careful inspection shows a strong spatial inhomogeneity of Δ (Fig. 4c) that relies nonmonotonically on the $E_F$ - $E_i$ value (Figs. 4d and 4e). The superconducting gap emerges at a certain threshold of $E_F$ - $E_i$ (i.e. the electron doping level), increases and then decreases in magnitude as the local doping level is increased further. This describes a primary source of the weak correlation between them (Fig. 4b). Such a nonmonotonic variation of Δ with the local doping level is derived from one sample by taking advantage of the dopant-induced nanoscale electronic inhomogeneity and differs from the nodal *d*-wave gap behavior previously reported on the charge reservoir planes [2,8], which declines linearly with the chemical doping. Instead, it bears a resemblance to the dome-shaped ($T_c$ *versus* doping) superconducting phase diagram of both electron- and hole-doped cuprates [1,2,28,29]. This indicates that the observed nodeless gaps on the $CuO_2$ plane are intimately linked to the superconducting properties ($T_c$) of cuprates.

**Conclusion**

We have arrived at our key findings of superconducting $CuO_2$ planes in SNCO, namely the nodeless electron pairing and spectroscopic evidence for the lattice vibrational modes in the superconducting domains. Although the $E_F - E_i$ and Δ vary significantly from domain to domain, the superconducting gaps are always fully opened on the Fermi surface (Figs. 3, 4d, S1 and S2 in the online supplementary material). A statistical average of Δ from > 3400 dI/dV spectra over many superconducting domains of all nine SNCO samples we studied (Fig. S1 in the online supplementary material) yields a value of 27 ± 8 meV. This mean gap size and the maximum gap $Δ_{max}$ ~ 40 meV observed in Fig. 4c stand out to be the highest of records in all electron-doped cuprate superconductors [19,20,28], and are comparable to those reported in the hole-doped counterparts [29]. Under this context, equivalently high-$T_c$ superconductivity might be potentially realized in the electron-doped infinite-layer cuprates once the inherent sample inhomogeneity is best minimized [30].

Our direct observation of nodeless superconductivity in the electron-doped cuprates of SNCO differs from the prior STM probe of a nodal *d*-wave gap function on the charge reservoir planes of various hole-doped cuprates [8]. Although it is tempting to examine the effect of an antiferromagnetic order on the nodeless energy gaps [20,31], our results exhibit much better consistency with those on the superconducting $CuO_2$ planes [9,32,33]. It thus becomes highly desirable to revisit the role of charge reservoir layers during the tunneling measurements of cuprates, and to testify whether the nodeless electron pairing is generic to the $CuO_2$ planes of the copper oxide superconductors. Combined with the simultaneous measurements of lattice vibrational modes, which are indiscernible from tunneling spectra of the non-superconducting domains due to the vanishing paired electrons there (Fig. 4d), our results agree with a phonon-mediated *s*-wave pairing state in SNCO. However, two cautions are yet taken to simply explain the findings from the conventional wisdom of Bardeen-Cooper-Schrieffer (BCS) theory. One is the gap-to-$T_c$ ratio $2Δ/k_BT_c$ ~ 14 (Fig. 1d) that largely exceeds the weak-coupling BCS value of 3.53. The other one relates to the anomalous dome-shaped doping dependence of Δ, whereas the BCS theory predict no obvious dependence of Δ on doping. These unconventional features, which have been observed in fulleride superconductors [34] and monolayer FeSe films grown on the $SrTiO_3$(001) substrates as well [35-37], go beyond the weak-coupling BCS picture. They, however, do not necessarily violate a phonon-mediated superconducting state with the local nonretarded pairs [38]. From this point of view, our results demonstrate the vital significance of electron-lattice interaction in the superconductivity of infinite-layer cuprates [39]. A further measurement of the oxygen isotope effects on Δ and Ω helps understand the role of phonons in the observed nodeless superconductivity.

**Materials and Methods**

**Sample growth.** High-quality $Sr_{1-x}Nd_xCuO_2$ (0.008 < x < 0.110) thin films were epitaxially grown in an ozone-assisted molecular beam epitaxy (O-MBE) chamber that contains a quartz crystal microbalance (QCM, Inficon SQM160H) for precise flux calibration. Atomically flat $SrTiO_3$(001) substrates with different Nb doping levels of 0.05 wt% and 0.5wt% were heated to 1200°C under ultrahigh vacuum (UHV) conditions for 20 minutes to acquire a $TiO_2$ terminated surface. The epitaxial $Sr_{1-x}Nd_xCuO_2$ films for STM and transport measurements were prepared on the 0.5 wt% and 0.05 wt% Nb-doped $SrTiO_3$ substrates, respectively. As oxidant, the distilled ozone flux was injected from a home-built ozone system into the O-MBE chamber by a nozzle, ~ 40 mm away from the substrates. All samples were grown by co-evaporating high-purity metal sources (Nd, Sr and Cu) from standard Knudsen cells under an ozone beam flux of ~ $1.1 \times 10^{-5}$ Torr and at an optimized substrate temperature $T_{sub}$ of 550°C [17]. The lower $T_{sub}$ (< 500°C) was revealed to degrade severely the sample quality of SNCO, while the higher ones (> 610°C) result into another competing orthorhombic phase. After growth, the films were annealed in UHV at the identical $T_{sub}$ for 0.5 hour and then cooled down to room temperature.

Prior to every film growth, we calibrated the beam flux of metal sources in sequence, to ensure the stoichiometry of SNCO films. The growth rate was kept at ~ 0.4 unit cell per minute. The doping level *x* is nominally deduced by *in-situ* QCM by calculating the flux ratio between the Nd and Cu sources, with an experimental uncertainty of approximately 0.5%. At the same time, the satellite peaks (Kiessig fringes) in the x-ray diffraction (XRD) spectra also allow us to estimate the film thickness [16,17] that agrees nicely with the nominal one deduced by the QCM-measured flux of Cu and growth duration.

***In-situ* STM measurements.** All STM measurements were performed in a Unisoku USM 1300S $^3$He system, which is connected to the O-MBE chamber, at a constant temperature of 4.8 K, unless otherwise specified. The system pressure is lower than $1.0 \times 10^{-10}$ Torr. Polycrystalline PtIr tips were cleaned via *e*-beam bombardment and calibrated on MBE-prepared Ag/Si(111) prior to the STM measurements. The STM topographies were acquired in a constant current mode with the voltage applied on the sample. The differential conductance dI/dV spectra and maps were measured by using a standard lock-in technique with a small bias modulation at 937 Hz. The system grounding and shielding have been optimized to increase the stability and spectroscopic energy resolution ($\sim$ 1.0 meV) of our STM apparatus.

**Transport measurements.** After *in-situ* STM characterization and *ex-situ* XRD measurements, the transport measurements were carried out in a standard physical property measurement system (PPMS, Quantum Design). Freshly-cut indium dots were cold pressed onto the samples as contacts. The resistivity was measured in a four-terminal configuration by a standard lock-in technique with a typical excitation current of 1 μA at 13 Hz.

**Supplementary data**

Supplementary data are available at NSR online.

**Funding**

The work was financially supported by the National Natural Science Foundation of China (Grants No. 51788104, No. 11634007, and No. 11774192, No. 11790311), and the Ministry of Science and Technology of China (2018YFA0305603, 2017YFA0304600 and 2017YFA0302902).

**Author contributions**

C.L.S., X.C.M. and Q.K.X. conceived the project. J.Q.F., X.Q.Y. and F.J.C. synthesized the thin film samples and performed the STM experiments with assistance from X.P.H.. H.W. and D.Z. carried out the transport measurements. J.Q.F. analyzed the data with assistance from X.Q.Y., F.J.C., R.F.W. and X.B.M.. C.L.S. and J.Q.F. wrote the manuscript with input from D.Z., X.C.M. and Q.K.X.. All authors discussed the results and commented on the manuscript.

**Competing interests:** None declared

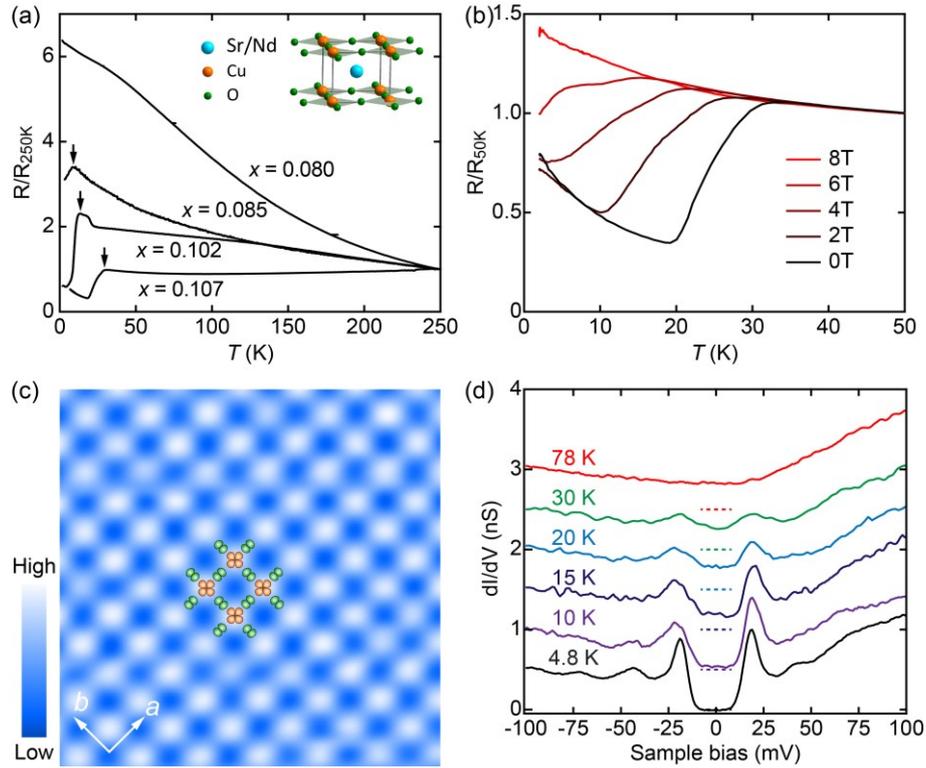

**Figure 1.** Characterizations of SNCO epitaxial films. (a) Temperature dependence of electrical resistivity, normalized to the value at 250 K, in electron-doped $Sr_{1-x}Nd_xCuO_2$ ($0.080 \leq x \leq 0.107$) cuprate films, with a nominal thickness of ~ 13 nm. Arrows denote the onset temperatures of superconductivity. Inset shows the schematic crystal structure of SNCO. (b) Electrical resistivity *versus* temperature of $Sr_{0.893}Nd_{0.107}CuO_2$ measured under different magnetic fields, normalized to the value at 50 K for clarify. (c) Atom-resolution STM topography (3.6 nm × 3.6 nm, $V$ = -1.5 V, $I$ = 20 pA) of $x \sim 0.100$ SNCO film. The bright spots denote the Cu atoms at the top layer. A single $CuO_2$ plaquette with Cu 3d (orange) and O 2p (green) orbitals is shown. (d) Temperature dependence of differential conductance dI/dV spectra on the superconducting $CuO_2$ plane. Setpoint: $V$ = -200 mV and $I$ = 100 pA.

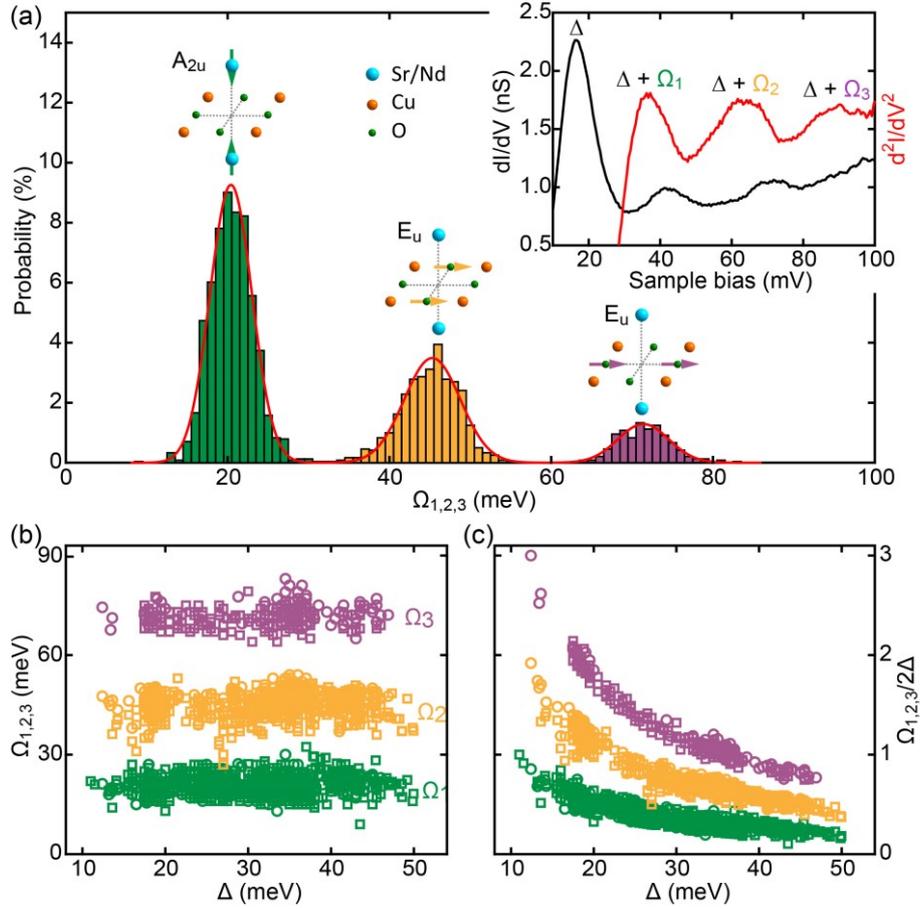

**Figure 2.** Lattice vibrational modes. (a) Histogram of measured bosonic mode energies Ω from a sequence of dI/dV spectra in nine similar SNCO samples with x ∼ 0.10. By fitting the data to a multipeak Gaussian function, three distinct bosonic modes at $\Omega_1 = 20 \pm 3$ meV, $\Omega_2 = 45 \pm 4$ meV and $\Omega_3 = 72 \pm 3$ meV are obtained, with their energies close to those of the external (green arrows), bending (yellow arrows) and stretching (purple arrows) phonons of copper oxides [22, 23]. Here the statistical errors of $\Omega_{1,2,3}$ indicate the full width at half maximum of the corresponding Gaussian peaks. Inserted in the top right corner are a representative superconducting spectrum (setpoint: *V* = -200 mV and *I* = 100 pA) and its derivative in the empty states, from which $\Omega_{1,2,3}$ can be readily extracted. (b) and (c) Variation of the bosonic mode energies $\Omega_{1,2,3}$ and $\Omega_{1,2,3}/2\Delta$ with the spatially varying gap magnitude Δ. Two different symbols of circles and squares denote data extracted from the occupied and empty states, respectively.

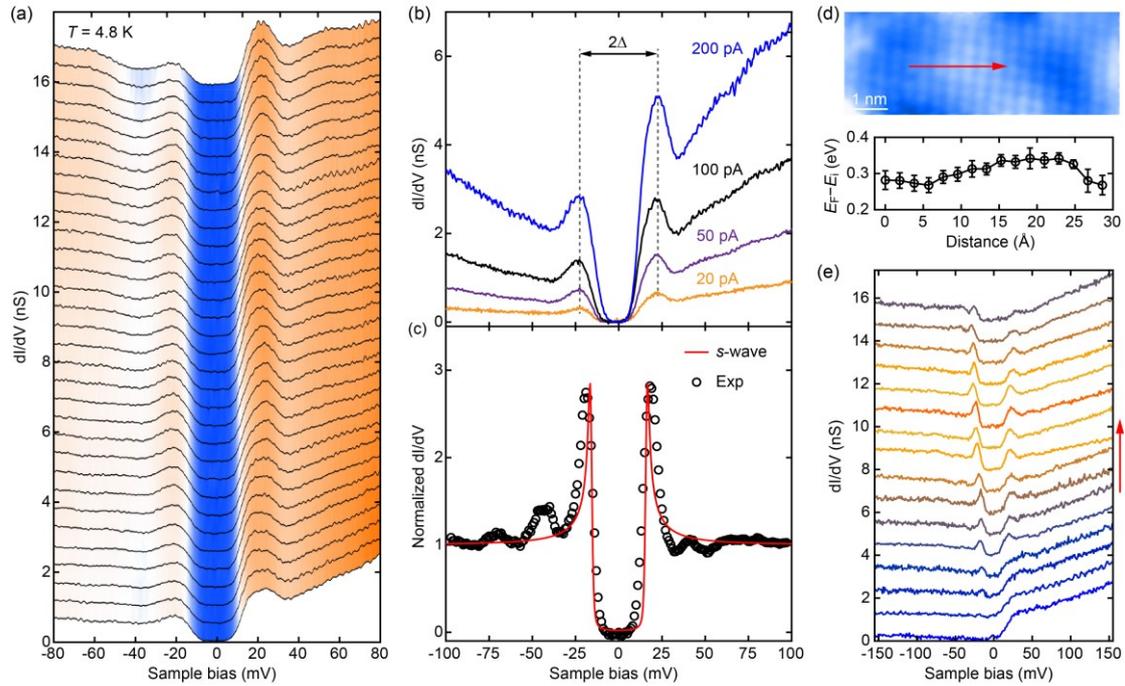

**Figure 3.** Tunneling spectra and nodeless superconductivity on $CuO_2$. (a) Line-cut dI/dV spectra taken at equal separations (0.05 nm) in one superconducting domain. Setpoint: $V$ = -100 mV and $I$ = 100 pA. (b) Tip-to-sample distance dependence of dI/dV spectra at a specific position. The tunneling current $I$ is changed from 20 pA (large distance, bottom curve) to 200 pA (small distance, top curve) at a constantly stabilized $V$ = -100 mV. (c) Normalized dI/dV spectrum with pronounced coherence peaks at 4.8 K (circles) and its best fit (red curve) to a single s-wave superconducting gap with Δ = 17 meV. The normalization was performed by dividing the raw dI/dV spectrum by its background, which was extracted from a quadratic fit to the conductance for $|V|$ > 50 mV. (d) An 8 nm × 3 nm STM topography ($V$ = -1.0 V, $I$ = 20 pA) showing nanoscale phase separation between superconducting (bright) and insulating (dark) domains. The bottom panel shows $E_F$ shifts relative to $E_i$ along the red arrow. Here a larger $E_F - E_i$ means heavier electron doping. (e) A series of dI/dV spectra acquired along the red arrow in d, plotted from bottom to top. Setpoint: $V$ = 200 mV and $I$ = 100 pA.

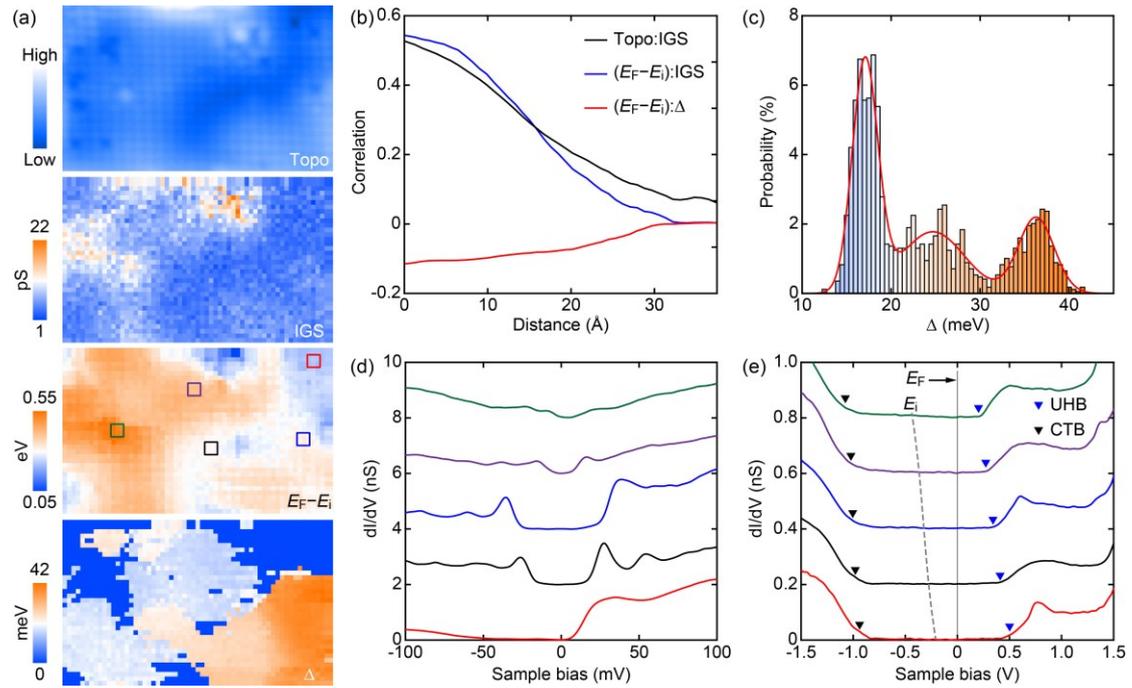

**Figure 4.** Spectroscopic mapping of nanoscale electronic phase separation. (a) Topography (10.5 nm × 6.2 nm, $V$ = -0.8 V, $I$ = 20 pA), spatial maps of IGS, $E_F - E_i$ and $\Delta$ extracted from a grid (61 pixels × 36 pixels) spectroscopic data over the same field of view. The STM tip is stabilized at $V$ = -1.6 V, $I$ = 100 pA and $V$ = -200 mV, $I$ = 100 pA to measure the dI/dV spectra in the wide and narrow voltage ranges, respectively. (b) Angle-averaged cross-correlations of IGS with the STM topography (black) and $E_F - E_i$ (blue), as well as cross-correlation between the $\Delta$ and $E_F - E_i$ (red). (c) Histogram of the superconducting gap $\Delta$. Three discrete peaks from the multipeak Gaussian fit (red) arise from various superconducting domains with different doping levels. (d) and (e) Spatially-averaged dI/dV spectra on the square-marked regions in the second lowest panel of (a), color coded to match with each other, measured in both narrow and wide energy scale ranges, respectively. The dashed line tracks the evolution of $E_i$ that evenly separates between the CTB (black triangles) and upper-Hubbard band (UHB, blue triangles) of the $CuO_2$ plane, while the gray solid one denotes $E_F$.

*Supplementary Material for*

# Direct observation of nodeless superconductivity and phonon modes in electron-doped copper oxide $Sr_{1-x}Nd_xCuO_2$

Jia-Qi Fan[1], Xue-Qing Yu[1], Fang-Jun Cheng[1], Heng Wang[1], Ruifeng Wang[1], Xiaobing Ma[1], Xiao-Peng Hu[1], Ding Zhang[1,2,3,4], Xu-Cun Ma[1,2 †], Qi-Kun Xue[1,2,3,5 †], Can-Li Song[1,2 †]

[1]State Key Laboratory of Low-Dimensional Quantum Physics, Department of Physics, Tsinghua University, Beijing 100084, China

[2]Frontier Science Center for Quantum Information, Beijing 100084, China

[3]Beijing Academy of Quantum Information Sciences, Beijing 100193, China

[4]RIKEN Center for Emergent Matter Science (CEMS), Wako, Saitama 351-0198, Japan

[5]Southern University of Science and Technology, Shenzhen 518055, China

†To whom correspondence should be addressed. Email: clsong07@mail.tsinghua.edu.cn, xucunma@mail.tsinghua.edu.cn, qkxue@mail.tsinghua.edu.cn

**This supplement includes:**

Fig. S1 to S5

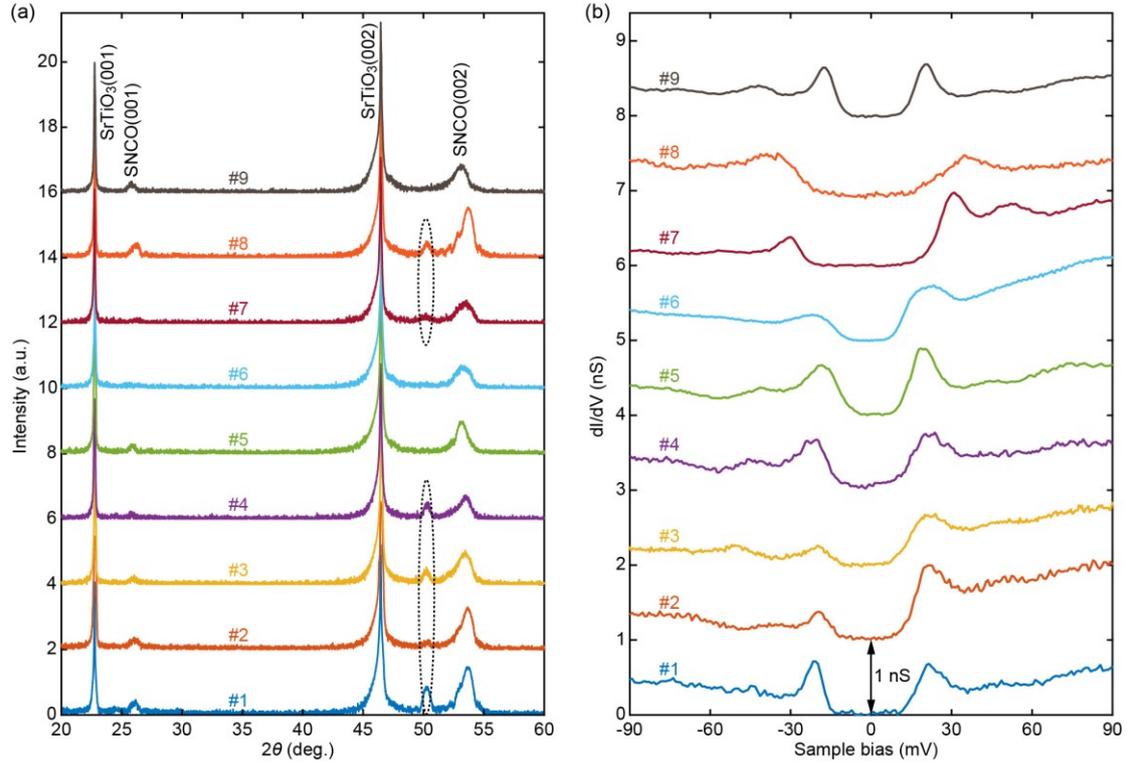

**Figure. S1.** Data reproducibility and robust nodeless superconductivity. (a) XRD spectra of nine similar SNCO samples (from #1 to #9) with a nominal doping $x \sim 0.100$ measured by using the monochromatic Cu $K_{\alpha 1}$ radiation with a wavelength of $\lambda = 0.15406$ nm. The experimental error of actual Nd dopant concentration results into a tiny trace of hole-doped SNCO (marked by the dashed ovals), caused by the appreciable intake of apical oxygens [16,17]. In this study, only the electron-doped SNCO regions have been explored with interest. (b) Tunneling dI/dV spectra consistently showing full superconducting gaps in various SNCO samples, color-coded to match the XRD spectra for the same sample in (a). Every curve corresponds to spatially averaged dI/dV spectrum in one typically superconducting domain. For clarity, the curves are vertically offset by 1 nS. The tunneling junction was stabilized at $I = 100$ pA and $V = -200$ mV, except for #1 ($V = 200$ mV), #4 ($V = -150$ mV), #5 ($V = -250$ mV) and #6 ($V = -100$ mV).

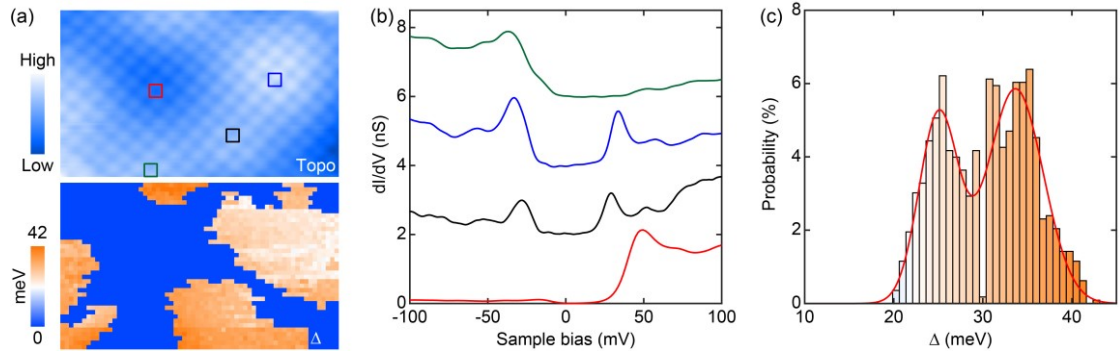

**Figure. S2.** Spectroscopic mapping of electronic phase separation and spatial inhomogeneity in Δ. (a) STM topography (7.0 nm × 4.2 nm, *V* = -0.8 V, *I* = 20 pA) and Δ map extracted from a grid (64 pixels × 38 pixels) spectroscopic data over the same field of view. The blue regions exhibit no spectroscopic sign of superconductivity, for which we assign Δ as zero. (b) Spatially-averaged tunneling dI/dV spectra on the square-marked regions in (a), color coded to match with each other. Setpoint: *V* = -200 mV and *I* = 100 pA. (c) Histogram of the measured superconducting gaps. Two discrete peaks of Δ from the multipeak Gaussian fit (red line) arise from different superconducting domains with varying doping levels.

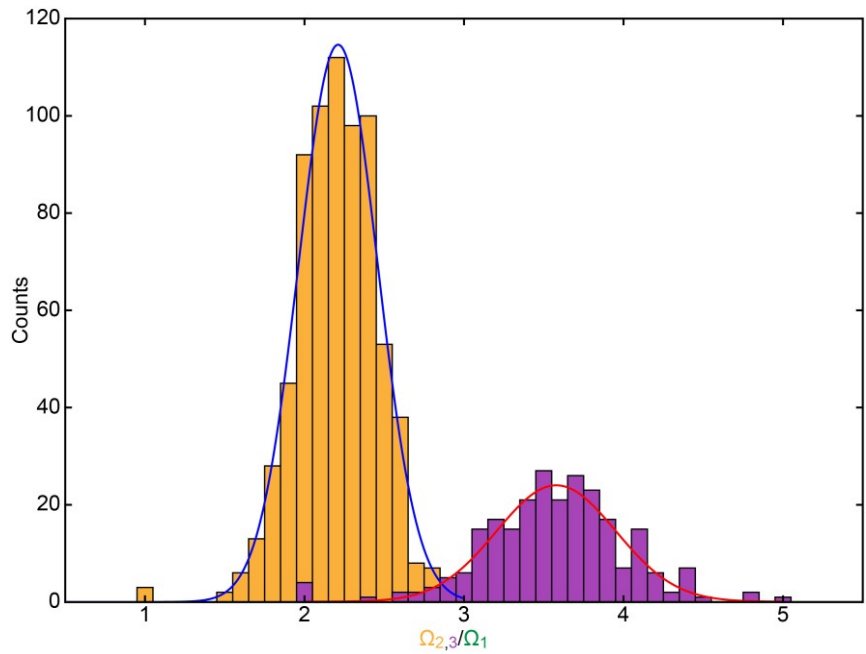

**Figure. S3.** Histograms of the mode energy ratio of $\Omega_2/\Omega_1$ (orange) and $\Omega_3/\Omega_1$ (purple). Each data has been carefully extracted from one identical spectrum with discernible $\Omega_2/\Omega_1$ and/or $\Omega_3/\Omega_1$. Blue and red solid lines denote Gaussian fits used to determine the ratio $\Omega_2/\Omega_1$ = 2.21 ± 0.01 and $\Omega_3/\Omega_1$ = 3.58 ± 0.02, respectively. The large deviations of $\Omega_2/\Omega_1$ and $\Omega_3/\Omega_1$ from integers exclude identifications of $\Omega_2$ and $\Omega_3$ as multiples of the same mode $\Omega_1$.

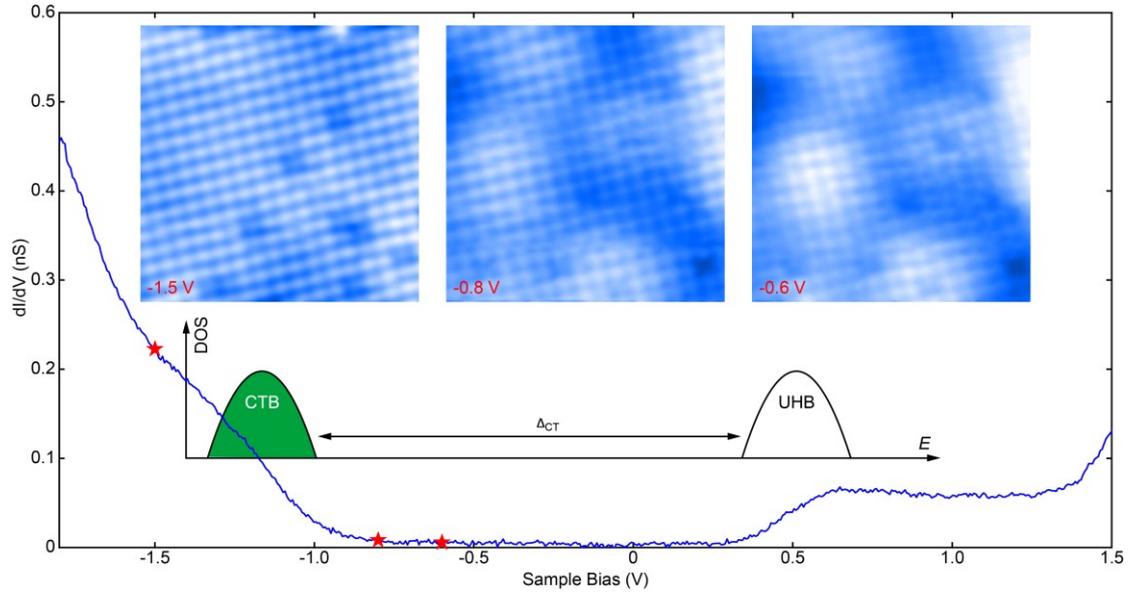

**Figure. S4.** Bias-dependent STM imaging contrast of CuO$_2$. Spatially-averaged dI/dV (blue curve) spectrum in one SNCO epitaxial film of $x \sim 0.100$, stabilized at $V$ = -1.8 V, $I$ = 100 pA. Inserted are bias-dependent STM topographies (6.2 nm × 6.2 nm, $I$ = 20 pA) in the same field of view and the schematic band structure of pristine cuprate, only showing the UHB (unfilled) and CTB (green). The three red stars from left to right mark the sample biases of $V$ = -1.5 eV, -0.8 eV and -0.6 eV applied to acquire the atomically-resolved STM topographic images in the top panel. Apparently, a nanometer-scale STM contrast becomes dominant for sample biases located within the charge-transfer gap ($\Delta_{CT}$), while the STM topography measured at -1.5 V exhibit no such contrast. This strongly hints at an electronic origin of the corresponding STM imaging contrast, primarily caused by a local doping variation of trivalent neodymium.

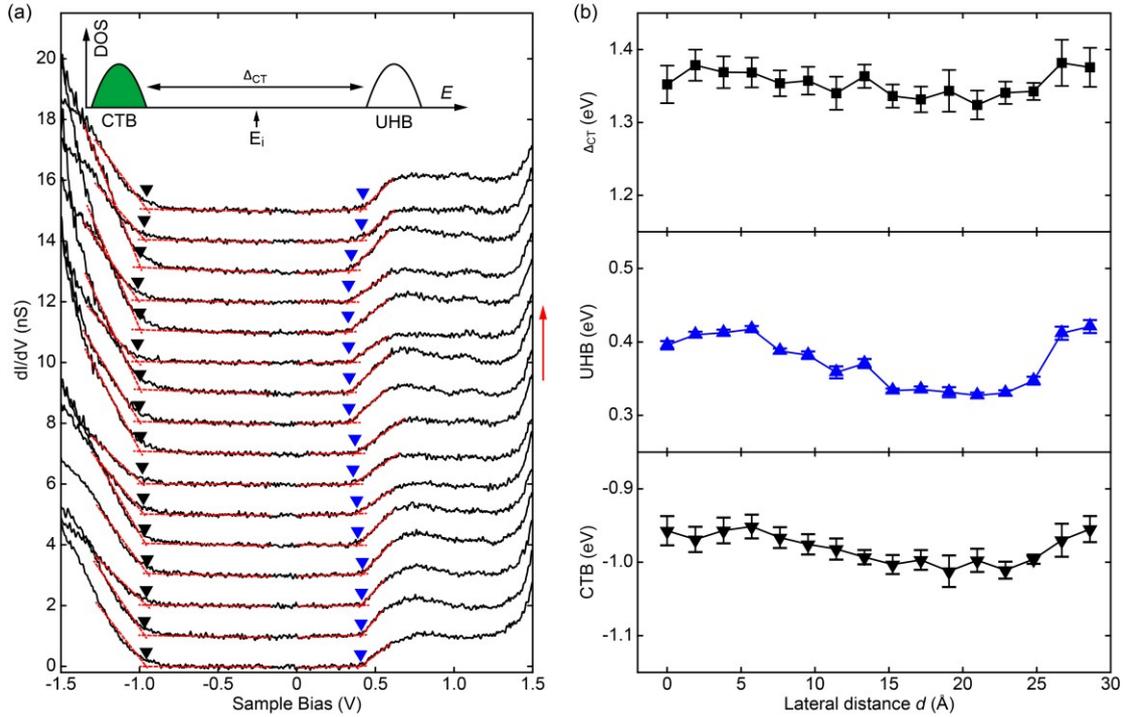

**Figure. S5.** Dopant-induced nanoscale electronic inhomogeneity and Mott parameters in SNCO. (a) Tunneling dI/dV spectra measured at equal separation (∼ 0.18 nm) along the red arrow in Fig. 3(d), illustrating the spatial variations of the CTB (black triangles) and UHB (blue triangles) onsets on the $CuO_2$ plane of SNCO. The spectra were taken by stabilizing the setpoint at $V$ = 1.5 V and $I$ = 100 pA. Red dashed lines indicate the linear fits to the electronic DOS justly below and above CTB/UHB, and the points of interaction are defined as the CTB/UHB onset energies. Inserted are schematic energy bands of cuprates, showing the CTB (green), UHB (unfilled) and the midgap energy $E_i$ (i.e. the center of charge transfer gap, see the black arrow). (b) Determined charge transfer gap $\Delta_{CT}$ (top panel), onset energies of UHB (middle panel) and CTB (bottom panel) as a function of the measured position $d$ from bottom to top along the red arrow in (a). The error bars arise from the uncertainties of linear fits to the electronic DOS near the band edges for calculating the UHB and CTB onsets.